\documentclass{elsart}
\usepackage{graphicx,amssymb}
\usepackage{amsmath,wrapfig,fancybox,bm}
\usepackage{fancybox}
\usepackage{bm}
\usepackage{color}
\begin{document}

\newtheorem{notation}[thm]{Notation}
\newtheorem{note2}[thm]{Note}
\journal{xxx}
\begin{frontmatter}
\title{A classification of phenomena in a two double-slit experiment in terms of symmetries of permutations of particle labels and attributes}
\author{Genta Ito}
\ead{cxq02365@gmail.com}
\address{Maruo Lab., 500 El Camino Real \# 302, Burlingame, CA 94010, United States}
\begin{abstract}
Permutations of particle labels are usually used to illustrate the relationship between classical and quantum statistics.  We use permutations of attributes/properties of particles to express properties of waves.  We express events of the well-known two-double-slit experiment (in regimes that are appropriate to quantum interference and classical interference) in terms of symmetries of permutations of particle labels and attributes.  We also obtain a new system called residual interference, and suggest an experiment to detect it.
\end{abstract}
\end{frontmatter}

\section{Introduction}\label{SecIntro_p1}
A distinction between two objects is made by indicating a difference in the value of at least one of their attributes.  For example, consider the attributes {\it purpose} and {\it composition}. Both a pencil and a ballpoint pen are used for the purpose of writing, so they have the same value of the attribute {\it purpose}; however, a pencil contains graphite but not ink, and a ballpoint pen contains ink but not graphite. Thus we can distinguish a pencil from a pen by their different values of the attribute {\it composition}.  

\begin{figure}[h]
\begin{center}
\includegraphics[width=50.4mm,height=7.2mm]{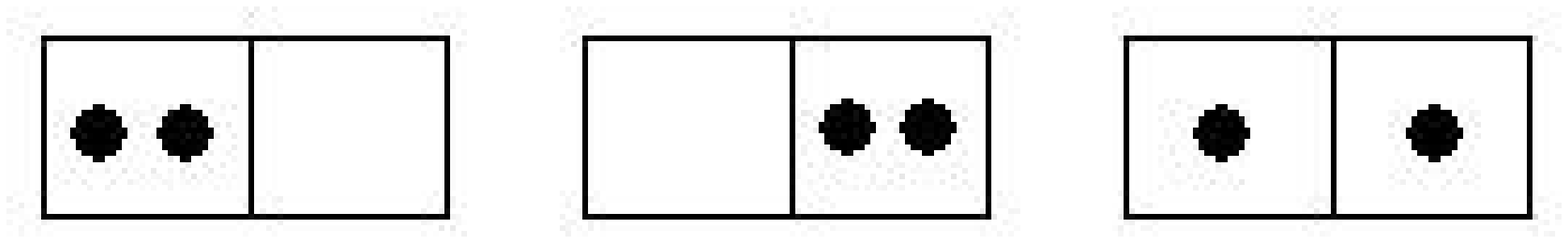}
\end{center}
\caption{}\label{FigStatistics_p1}\end{figure}

Consider the distribution of two indistinguishable particles over two boxes, which represent a physical attribute such as spin (see Fig.~\ref{FigStatistics_p1}).

Let $L$ and $R$ denote the left and right boxes (corresponding to ``left'' and ``right'' as values of the attribute), respectively, in Fig.~\ref{FigStatistics_p1}, and express the event that particle 1 is in box $L$ as $L(1)$, etc.  It is important to note that $L$ denotes the value of an attribute but $L(1)$ denotes an event.  

Also let us say that an attribute is {\it elementary} if it applies to only {\it one} object,  and that a {\it combined} attribute is one that consists of a pair of distinct elementary attributes.  Then $L$ and $R$ are the elementary attributes, and $LR$ is the combined attribute. Similarly, let us define an {\it elementary} event as one that specifies an elementary attribute and a particle label to go with it (e.g., $L(1)$ is an elementary event), and a {\it combined} event as one that specifies a pair of distinct elementary attributes and a pair of distinct particle labels attached to those attributes (e.g., $L(1)R(2)$ and $L(2)R(1)$ are combined events). 

Permutations of particle labels can be used to illustrate the difference between classical statistics and quantum statistics.  In classical statistics, a permutation of the particle labels in the third arrangement (where one particle is in the left box and the other is in the right box) is counted as giving rise to a different arrangement---which implies that, although the particles are indistinguishable, they can be regarded as individuals.  That is, in classical statistics, the combined events, $L(1)R(2)$ and $L(2)R(1)$, are treated as different events, so that the set of all possible arrangements is as follows:
\[\{L(1)L(2),\ R(1)R(2),\ L(1)R(2),\ L(2)R(1)\}\]
In quantum statistics, a permutation of the labels of the particles in the third arrangement is not regarded as giving rise to a new arrangement, so quantum particles cannot be regarded as individuals in the same sense as classical particles. That is, in quantum statistics, $L(1)R(2)$ and $L(2)R(1)$ are treated as a single event, which is interpreted as the label-{\it symmetric} event $L(1)R(2)+L(2)R(1)$ in Bose--Einstein statistics and the label-{\it antisymmetric} event $L(1)R(2)-L(2)R(1)$ in Fermi--Dirac statistics. An event is {\it label symmetric} (resp. {\it antisymmetry}) if it remains unchanged (resp. changed) under any permutation of particle labels. It should be noted that $L(1)R(2)+L(2)R(1)$ and $L(1)R(2)-L(2)R(1)$ do not indicate superposition of $L(1)R(2)$ and $L(2)R(1)$~\cite{Teller2}.

Thus the set of arrangements for Bose--Einstein statistics is as follows:
\[\{L(1)L(2),\,R(1)R(2),\,L(1)R(2)+L(2)R(1)\}\]
If our purpose were to examine the difference between Bose--Einstein statistics and Fermi--Dirac statistics, we would have to focus on the {\it difference} between label {\it symmetry} and label {\it antisymmetry}.  However, our sole purpose here is to examine the difference between classical and quantum statistics, so we will consider only what Bose--Einstein statistics and Fermi--Dirac statistics have in common.  Thus what follows will apply to Bose--Einstein statistics, though it could just as well be applied to Fermi--Dirac statistics.  

Just as permutations of particle labels illustrate the difference between classical statistics and quantum statistics, permutations of the values of an attribute (such as $L$ and $R$) can be regarded as representing indistinguishability of some kind.  Thus we can translate the wave property of a quantum particle into a permutation of the values of some attribute of that particle.  In a double-slit experiment, for example, we cannot determine whether a particle goes through slit \textbf{A} or slit \textbf{B}, and we can express that by saying that the event is symmetric with respect to interchange of \textbf{A} and \textbf{B}. 

Although the heart of quantum reality is typically discussed in terms of the one-double-slit experiment~\cite{FeynmanLecture}, the experiment is regarded as a pedagogical subject from the viewpoint of modern physics. However, it remains unclear what kind of uncertainty relation is involved there. The incompatibility between distinguishing the particle path and viewing the interference pattern is explained as a consequence of the uncertainty relation in some textbooks~\cite{FeynmanLecture,MessiahText}. Tanimura~\cite{Tanimura1} formulated a quantitative relation of the incompatibility, and showed that the uncertainty relation relevant to the double-slit experiment is not the Ozawa-type uncertainty relation~\cite{Ozawa1} but the Kennard-type uncertainty relation~\cite{Kennard} of the position and the momentum of the double-slit wall.

In classical physics, {\it location in spacetime}  can be treated as an ultimate attribute, because different objects cannot occupy the same point in spacetime~\cite{Post,French,Teller1,Huggett}.  However, this property of classical objects, which is called impenetrability, does not hold of waves---or of certain entities that involve functional identity.  Objects are individual because of impenetrability, but things are not, and therefore there is no effective measure of distinction for things that have the same properties.  Also, impenetrability does not hold of objects in SQM, so the hard problem of encoding individuality into an elementary measurement reduces to the question of what it means for states to be indistinguishable and yet individual. Thus, standard quantum theory (SQT) does not permit us to assume that matter consists of particles pursuing definite tracks in spacetime, and the complete description of a system is provided by its wavefunction.  However, Bohmian quantum theory (BQT) reintroduced the impenetrability (i.e., the particle position in the classical sense)~\cite{Bohm1,Bohm2}. 

Ghose suggested a one-double-slit experiment with specially entangled two-bosonic particles in order to study the difference between SQT and BQT~\cite{Ghose1,Ghose2}. Both theories are equivalent where dynamical variables averaged over a Gibbs ensemble of Bohmian trajectories; however BQM makes explicit predictions about the particles which are emitted separately (i.e., emitted in clearly separated short intervals of time).

Golshani and Akhavan suggested a two double-slit experiment (rather than the one-double-slit experiment) to study the difference between SQM and BQM. Both theories lead the same final interference pattern; however BQM makes explicit predictions about each pair of particles which are emitted separately.  In the experiment, the deviation of the source from its geometrically symmetric location is unnecessary to obtain a different interference pattern from SQT~\cite{GolshaniAkhavan2001}.

In this paper we will classify interference patterns in a two double-slit experiment in terms of symmetries of permutations of particle labels and attributes. In section~\ref{SecDoubleSlit_p1}, we will introduce some uncertainty in the coordinates of the point of emission, which is uncertainty of the source from its geometrically symmetric location, and then use SQT to obtain two interference patterns, classical interference (CI) and quantum interference (QI). In CI, particle paths are indistinguishable but the interference occurs on both local screens, whereas, in QI, particle paths are distinguishable but the interference does not occur on each local screen. In section~\ref{SecAttributesEventsSymmetriesClassification_p1}, the incompatibility between CI and QI, which is explained as a consequence of the (usual) uncertainty relation (in the vertical coordinate), will be explained in terms of permutations of particle labels and attributes.  Also, we will introduce a third system called residual interference (RI), which is also subject to the uncertainty relation in the vertical coordinate, but is separate from QI and CI. Furthermore, in section~\ref{SecRemarks_p1}, we will suggest an experiment to detect RI.

\section{Two-double-slit experiment}\label{SecDoubleSlit_p1}
\begin{figure}[h]
\begin{center}
\includegraphics[width=128.2mm,height=74.6mm]{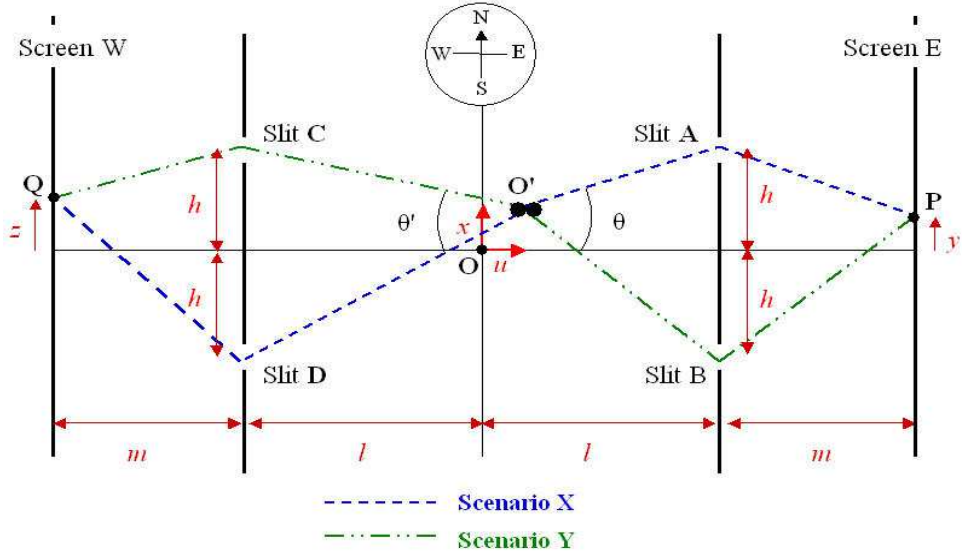}
\end{center}
\caption{Two-double-slit experiment}\label{FigTwoDoubleSlit_p1}
\end{figure}
Suppose that, as shown in Fig.~\ref{FigTwoDoubleSlit_p1}, a pair of photons with wavelength $\lambda$ and wavenumber $ k=2\pi/\lambda$ are emitted at the origin $\mathbf{O}$.  There is some uncertainty in the coordinates of the point of emission, so assume that the photons are actually emitted at some point $\mathbf{O}^\prime(u,x)$ near the origin, and that the horizontal and vertical coordinates of $\mathbf{O}^\prime$ (namely, $u$ and $x$) lie in the intervals $[u_{1},u_{2}]$ and $[x_{1},x_{2}]$, respectively.  Thus if the uncertainty $\Delta u$ in the horizontal coordinate of $\mathbf{O}^\prime$ (i.e., the uncertainty $u_{2}-u_{1}$) is 0, then $u=0$, and similarly for the uncertainty $\Delta x$ in the vertical coordinate.

Suppose, furthermore, that one photon passes through slit $\bm{\mathrm{A}}\left(l,h\right)$ and/or slit $\bm{\mathrm{B}}\left(l,-h\right)$, and reaches point $\bm{\mathrm{P}}(l+m,y)$ on screen E, and that the other photon passes through slit $\bm{\mathrm{C}}\left(-l,h\right)$ and/or slit $\bm{\mathrm{D}}\left(-l,-h\right)$, and reaches point $\bm{\mathrm{Q}}(-l-m,z)$ on screen W.  As seen in the figure, $\theta$ and $\theta^{\prime}$ denote angles $\mathbf{AO^\prime B}$ and $\mathbf{CO^\prime D}$, respectively. The reason for indicating the directions north (N), south (S), east (E), and west (W) will become clear later.

Let $\Psi(y,z)$ be the probability amplitude that one photon reaches point $\mathbf{P}$ and the other photon reaches point $\mathbf{Q}$. We now derive an approximate expression for $\Psi(y,z)$. Then

\begin{eqnarray}
\label{Prob1_p1}
\displaystyle \Psi(y,z)&\sim& \displaystyle \frac{1}{(u_{2}-u_{1})\cdot(x_{2}-x_{1})}\int_{u_{1}}^{u_{2}}\int_{x_{1}}^{x_{2}}\mathrm{cos}\frac{kL_{E}}{2}\cdot\mathrm{cos}\frac{kL_{W}}{2}\ dudx
\end{eqnarray}
where $L_E$ and $L_W$ are the path differences $\mathbf{\overline{O^{\prime}BP}}-\mathbf{\overline{O^{\prime}AP}}$ and $\mathbf{\overline{O^{\prime}DQ}}-\mathbf{\overline{O^{\prime}CQ}}$, respectively. We integrate over $u$ and $x$, in order to include the contributions of photons from all possible initial points $\bm{\mathrm{O}}^\prime(u,x)$ with $u_{1}\leq u\leq u_{2}$ and $x_{1}\leq x\leq x_{2}$.  

First, we assume that $l$ and $m$ are very large compared to $x,u,y,z$, and $h$, and that $x\ll h/2$.  Then the path difference $L_{E}$ can be approximated by $\displaystyle \frac{hx}{l}+\frac{hy}{m}$. 
Similarly, $L_{W}\approx W_{D}-W_{C}$. Hence the small horizontal displacement $u$ of point $\mathrm{O}^{\prime}$ from the origin does not appear in the final formulas for the path differences $L_{E}$ and $L_{W}$, since the effect of $u$ on each of them is negligible.  Thus the integral over $u$ in (\ref{Prob1_p1}) is just $(u_{2}-u_{1})$.

Since we assumed that $h$ is much smaller than $l$, and that $x$ is much smaller than both $h$ and $l$, we have $\tan(\theta/2)\approx\left(h/l\right)$ and $\tan\left(\theta/2\right)\approx\theta/2$;  hence $\theta\approx 2h/l$.  By analogous reasoning, $\theta^{\prime}\approx 2h/l$. Thus we can consider $\theta$ and $\theta^{\prime}$ to be independent of $x$.  

For the sake of convenience, suppose that $l=m$, and, in addition, that there is some positive number $d$ such that $x_{1}=-d/2$ and $x_{2}=d/2$.  This gives

\begin{eqnarray}
\nonumber
\displaystyle \Psi(y,z)&\sim& \displaystyle \frac{1}{d}\int_{-d/2}^{d/2}\mathrm{cos}\frac{k\theta(x+y)}{2}\cdot \mathrm{c}\mathrm{o}\mathrm{s}\frac{k\theta(x+z)}{2}dx\\
\label{Prob_p1}
&=&\displaystyle \frac{2}{k\theta d}\mathrm{sin}\frac{k\theta d}{2}\cdot\frac{1}{2}\mathrm{c}\mathrm{o}\mathrm{s}\frac{k\theta(y+z)}{2}+\frac{1}{2}\mathrm{c}\mathrm{o}\mathrm{s}\frac{k\theta(y-z)}{2}
\end{eqnarray}

Let us classify phenomena by the uncertainty $d$ ($=\Delta x$) in the vertical position.

\textbf{Case (i)}  $k\theta d\ll 1$

The uncertainty in the vertical coordinate of the point of emission is $d=\triangle x$.  Also, the photon momentum is $p=\hbar k$, so $|p_{x}|\le\hbar k$, where $p_{x}$ is the component of the photon momentum in the vertical direction.  From the Uncertainty Principle ($\Delta x\cdot\Delta p_{x}\sim\hbar$, where $\Delta p_{x}$ is the uncertainty in $p_{x}$) and the inequality $k\theta d\ll 1$, we obtain $\triangle p_{x}/|p_{x}|\gg\theta$ 
if $|p_{x}|>0$.  The fact that $\triangle x$ is small in this case means that $\Delta p_{x}$ is large, and the inequality $\triangle p_{x}/|p_{x}|$ implies that one photon can pass through slits $\bm{\mathrm{A}}$ and $\bm{\mathrm{B}}$ simultaneously---a condition that can be described by saying that one photon {\it illuminates} the two slits simultaneously.  Also, another photon illuminates slits $\bm{\mathrm{C}}$ and $\bm{\mathrm{D}}$ simultaneously.  Moreover, it follows from the inequality $k\theta d\ll 1$ that $(2/k\theta d)\cdot\mathrm{sin}(k\theta d/2)\approx 1$, so (\ref{Prob_p1}) reduces to

\begin{equation}\label{CIWaveFunction_p1}
\displaystyle \Psi(y,z)\sim\frac{1}{2}(\mathrm{cos}\frac{k\theta(y+z)}{2}+\mathrm{cos}\frac{k\theta(y-z)}{2})=\mathrm{cos}\frac{k\theta y}{2}\cdot\mathrm{cos}\frac{k\theta z}{2}
\end{equation}

This means that there is classical interference (CI), because the interference on either screen is independent of that on the other screen.  The interference on screen E (which we will refer to as scenario {\bf E}) arises from the two slits on the right side of the diagram ({\bf A} and {\bf B}), whereas the interference on screen W (scenario {\bf W}) arises from the two slits on the left side ({\bf C} and {\bf D}).  Thus in CI we can say that there are two types of symmetry occurring simultaneously: symmetry of permutation of attributes of scenario {\bf E} (i.e., permutation of slits {\bf A} and {\bf B}), and symmetry of permutation of attributes of scenario {\bf W} (i.e., permutation of slits {\bf C} and {\bf D}). We will denote these symmetries by $(A,B)$ and $(C,D)$, respectively. Note, however, that even though in CI one photon illuminates the two slits on the right (east) side simultaneously and another photon illuminates the two slits on the left (west) side simultaneously, there is no information as to how the photon on the east side reaches screen E (i.e., whether it passes through both slits or just one---and if the latter, \textit{which} slit: \textbf{A} or \textbf{B}), nor is there any information as to how the photon on the west side reaches screen W.

\textbf{Case (ii)}  $k\theta d\gg 1$

From the Uncertainty Principle and the inequality $k\theta d\gg 1$, it follows that $\triangle p_{x}/|p_{x}|\ll\theta$.  Since $\triangle x$ is large in this case, $\Delta p_{x}$ is small.  Thus the direction of the vertical component of the momentum of one of the two photons is opposite that of the other photon.  In this case, there are two possible scenarios, $\mathbf{X}$ and $\mathbf{Y}$, where in $\mathbf{X}$ one photon passes through slit $\bm{\mathrm{A}}$ and the other passes through slit $\bm{\mathrm{D}}$, and in $\mathbf{Y}$ one photon passes through slit $\bm{\mathrm{B}}$ and the other passes through slit $\bm{\mathrm{C}}$.  Furthermore, the inequality $\triangle p_{x}/|p_{x}|\ll\theta$ implies that just {\it one} of these two scenarios can occur; hence it cannot be the case that one photon illuminates slits $\bm{\mathrm{A}}$ and $\bm{\mathrm{B}}$ and another photon illuminates slits $\bm{\mathrm{C}}$ and $\bm{\mathrm{D}}$.  Since $k\theta d\gg 1$, the first term of (\ref{Prob_p1}) is negligible, so we obtain 

\begin{equation}\label{QIWaveFunction_p1}
\displaystyle \Psi(y,z)\sim\frac{1}{2}\mathrm{cos}\frac{k\theta(y-z)}{2}
\end{equation}

This result indicates the presence of quantum interference (QI), since (\ref{QIWaveFunction_p1}) does not factor into separate expressions for interference on screen E and interference on screen W. Thus the effect is equivalent to interference that takes place on a fictional ({\it third}) screen, with an interference pattern which exhibits the fact that in QI illumination of one slit on the east side is inextricably linked with illumination of one slit on the west side. Because of conservation of momentum in the vertical direction, either symmetry $(A,D)$ or symmetry $(B,C)$ is acting, but not both. These symmetries are permutations of attributes associated with scenario $\mathbf{X}$ and scenario $\mathbf{Y}$, respectively. 

We can distinguish scenario $\mathbf{X}$ from scenario $\mathbf{Y}$; however, we do not count the event in which ``photon 1 reaches point $\mathbf{P}$ and photon 2 reaches point $\mathbf{Q}$'' as different from the event in which ``photon 1 reaches point $\mathbf{Q}$ and photon 2 reaches point $\mathbf{P}$.''  In other words, there is symmetry of permutation of the particle labels $(1,2)$.  This corresponds to our discussion of events in section~\ref{SecIntro_p1}: in quantum statistics, we count the sum $L(1)R(2)+L(2)R(1)$ as one event, rather than as the two events $L(1)R(2)$ and $L(2)R(1)$.  Therefore, QI is symmetric with respect to particle labels.

In CI, the same type of interference occurs on both screens, but we can macroscopically distinguish the interference on one screen from that on the other.  Therefore, we can distinguish the interference on screen E from the interference on screen W by permutation of the particle labels $(1,2)$.  In other words, CI is not symmetric with respect to permutation of particle labels.  This corresponds to our discussion of events in section~\ref{SecIntro_p1}: in classical statistics, we count $L(1)R(2)$ and $L(2)R(1)$ as two different events. Therefore, CI is asymmetric with respect to particle labels.

The heart of quantum reality is typically discussed in terms of the one-double-slit experiment.  Even in CI, however, it makes sense to discuss interference in terms of the two-double-slit experiment, since interference occurs on both sides of such an experiment (east and west) and the interference pattern on each side is identical to that for the one-double-slit experiment.  With the two-double-slit experiment as a backdrop, we will use particle labels to classify interference patterns in both QI and CI.  Also, we will introduce a third system (RI), which is separate from both QI and CI.

\section{Attributes, events, symmetries, and classification}
\label{SecAttributesEventsSymmetriesClassification_p1}
Let us regard slit $\bm{\mathrm{A}}$ as an ``attribute'' $A$ of a photon, namely, that a photon passes through slit $\bm{\mathrm{A}}$, and let attributes $B,\ C$, and $D$ correspond to slits $\mathbf{B}, \mathbf{C}$, and $\mathbf{D}$, respectively.  Denote the event that photon 1 passes through slit $\bm{\mathrm{A}}$ as $A(1)$, akin to the notation $L(1)$ introduced in section~\ref{SecIntro_p1}.  $A,\ B,\ C$, and $D$ are the elementary attributes, and $AB,\ AC,\ AD,\ BC,\ BD$, and $CD$ are the combined attributes. 

Next, let us regard scenario $\bm{\mathrm{X}}$ as a combined attribute (the combination of the two attributes that naturally correspond to scenario {\bf X}, namely, $A$ and $D$), and denote it by $X$. Furthermore, let $X(1,2)$ and $X(2,1)$ denote the corresponding combined events (the two events in which one object has attribute $A$ and the other has attribute $D$: $A(1)D(2)$ and $A(2)D(1)$, respectively). The combined event $X(1,2)=A(1)D(2)$ is interpreted as that in which photon 1 passes through slit $\bm{\mathrm{A}}$ and photon 2 passes through slit {\bf D}.  Similarly, let $Y,\ E,\ W,\ N,\ S$ be the combined attributes defined as follows: $Y=BC$,\ $E=AB$,\ $W=CD$,\ $N=AC$,\ $S=BD$. (The last four were named $E,\ W,\ N$, and $S$ because of their natural correspondence with the directions east, west, north, and south, respectively, in Fig.~\ref{FigTwoDoubleSlit_p1}.)

The elementary and combined attributes and events that pertain to the {\it two}-double-slit experiment in section~\ref{SecDoubleSlit_p1} are summarized in Table~\ref{TableNotations2_p1}.
\begin{table}[htbp]\caption{Attributes and events pertinent to the two-double-slit experiment}
\footnotesize
\begin{tabular}{|c|c|c|}  \hline
&Elementary&Combined\\ \hline
Attribute&$A,B,C,D$&$X=AD,\ \ Y=BC,\ \ E=AB$\\
&&$W=CD,\ \ N=AC,\ \ S=BD$\\ \hline
Event&$A(1),B(1),C(1),D(1)$&$X(1,2)=A(1)D(2),\ \ X(2,1)=A(2)D(1)$\\
&$A(2),B(2),C(2),D(2)$&$Y(1,2)=B(1)C(2),\ \ Y(2,1)=B(2)C(1)$\\
&&$E(1,2)=A(1)B(2),\ \ E(2,1)=A(2)B(1)$\\
&&$W(1,2)=C(1)D(2),\ \ W(2,1)=C(2)D(1)$\\
&&$N(1,2)=A(1)C(2),\ \ N(2,1)=A(2)C(1)$\\
&&$S(1,2)=B(1)D(2),\ \ S(2,1)=B(2)D(1)$\\ \hline
\end{tabular}\label{TableNotations2_p1}\end{table}

In section~\ref{SecIntro_p1}, we defined addition of (combined) events that pertain to the example given there. Here, we extend that definition to combined events that pertain to the two-double-slit experiment:  The event $X(1,2)+X(2,1)=A(1)D(2)+A(2)D(1)$ is interpreted as that in which one photon passes through slit $\bm{\mathrm{A}}$ and the other photon passes through slit $\bm{\mathrm{D}}$. 

There is a difference between a (combined) attribute such as $X=AD$ and an event such as $X(1,2)+X(2,1)=A(1)D(2)+A(2)D(1)$.  The difference lies in the symmetries possessed by combined attributes and sums of combined events. The combined attribute $X=AD$ has only the symmetry of permutation of the elementary attributes $A$ and $D$, while the event $X(1,2)+X(2,1)$ has the symmetry of permutation of labels 1 and 2 in addition to the symmetry of permutation of the underlying elementary attributes.  Therefore, when we discuss the distinction between classical statistics and quantum statistics (which is determined by the presence or absence of symmetry of permutation of particle labels), we have to consider events, not just the underlying attributes.  As explained in section~\ref{SecIntro_p1}, the event $L(1)R(2)+L(2)R(1)$ is symmetric with respect to permutation of particle labels and applies to quantum statistics, whereas the events $L(1)R(2)$ and $L(2)R(1)$ do not have such symmetry and apply to classical statistics. 

In what follows, we make several observations and provide new definitions pertinent to classification of the systems QI, CI, and RI according to the combined attributes and the sums of combined events that apply to each. 

\begin{itemize}
\item  Each of the three systems (QI, CI, RI) has two scenarios.  For example, QI has scenarios {\bf X} and {\bf Y}, each of which is naturally associated with a combined attribute: {\bf X} is associated with $AD$, because we cannot distinguish a photon that passes through slit {\bf A} from one that passes through slit {\bf D}; and {\bf Y} is associated with $BC$, because we cannot distinguish a photon that passes through slit {\bf B} from one that passes through slit {\bf C}. Thus it makes sense to discuss {\bf X} and {\bf Y} in terms of the combined attributes $X=AD$ and $Y=BC$, respectively, and similarly for the scenarios related to CI and RI.

\item  QI is directly related to the combined attributes $X$ and $Y$, and CI is directly related to $E$ and $W$.  We have chosen to call the new system ``RI'' (which stands for {\it residual interference}), because we are assigning to it the two ``residual'' (leftover) combined attributes $N,\ S$ (those that apply to neither QI nor CI).  QI is a phenomenon which is observable as either scenario {\bf X} or scenario {\bf Y}, and CI is a phenomenon which is observable as interference on screens $\mathrm{E}$ and $\mathrm{W}$ simultaneously. RI is assumed to have symmetries $N$ and $S$, but it is not known if RI is an observable phenomenon.

\item  Define an {\it even} event as a sum of two distinct combined events that contain a total of either two or four (an even number of) distinct elementary attributes. For example, $X(2,1)+Y(2,1)$ ($=A(2)D(1)+B(2)C(1)$) is an even event, since it contains all four elementary attributes ($A,\ B,\ C$, and $D$). There are a total of 18 even events. As we will see, each of the three systems (QI, CI, and RI) is naturally associated with six of them. The even events can be categorized in several ways:

\begin{enumerate}
\item[a.]  If  at least one of the two terms in an even event is (a combined event which is) prohibited, that even event is termed an {\it anti}-event. An even event that is not an anti-event is termed a {\it regular} event. For example, $X(2,1)+Y(2,1)$ ($=A(2)D(1)+B(2)C(1)$) is a regular event, and $E(1,2)+Y(1,2)$ ($=A(1)B(2)+C(1)D(2)$) is an anti-event. It follows from the definition of an even event that if at least one of the terms is prohibited, then both terms are prohibited. Thus we could have defined a regular event as an even event in which neither term is prohibited, and an anti-event as an even event in which both terms are prohibited.

\item[b.]  Define an event to be {\it auto-symmetric} (AS) if it is an even event which is invariant under permutation of a single pair of distinct elementary attributes. For example, $S(1,2)+S(2,1)$ ($=B(1)D(2)+B(2)D(1)$) is auto-symmetric, because it is invariant under the permutation $(B,D)$ of its elementary attributes, which yields $D(1)B(2)+D(2)B(1)$. To see that this is the original event ($S(1,2)+S(2,1)$), first switch the two terms (which gives $D(2)B(1)+D(1)B(2)$), and then switch the two elementary events within each term (i.e., switch $D(2)$ with $B(1)$ in the first term, and $D(1)$ with $B(2)$ in the second term), which gives $B(1)D(2)+B(2)D(1)$ ($=S(1,2)+S(2,1)$).

\noindent
It can easily be shown that an event is AS if and only if it is an even event which is invariant under the permutation $(1,2)$ of the particle labels. An event with the latter property is said to be {\it label symmetric} (LS), so an event is AS if and only if it is LS. In addition, it can be shown that an event is AS if and only if it contains a total of two distinct elementary attributes. Thus the LS events are those that contain a total of two distinct elementary attributes, and the non-LS (NLS) events are those that contain a total of four. Furthermore, every LS event is invariant with respect to each of two permutations (separately): permutation of a single pair of distinct elementary attributes (e.g., the LS event $E(1,2)+E(2,1)$ (=$A(1)B(2)+A(2)B(1)$) is symmetric with respect to the permutation $(A,B)$), and the permutation $(1,2)$ of the particle labels.

\item[c.]  Two pairs of non-LS events can be generated from any LS event. One pair is found by taking the two elementary events with particle label 1 in the LS event, and replacing the elementary attributes in those elementary events with the two elementary attributes that do not occur in the LS event. For example, the elementary attributes that do not occur in the LS event $W(1,2)+W(2,1)$ ($=C(1)D(2)+C(2)D(1)$) are $A$ and $B$. To obtain one non-LS event from $W(1,2)+W(2,1)$, replace the elementary attributes $C$ and $D$ in the elementary events $C(1)$ and $D(1)$ (the elementary events that have particle label 1) with the elementary attributes $A$ and $B$ (the elementary attributes that do not occur in $W(1,2)+W(2,1)$). If we replace the $C$ in $C(1)$ with $A$, and the $D$ in $D(1)$ with $B$, we obtain the non-LS event $A(1)D(2)+C(2)B(1)$; to write this as a sum of two combined events, switch the elementary events in the term $C(2)B(1)$, which yields $A(1)D(2)+B(1)C(2)$ ($=X(1,2)+Y(1,2)$). If instead we replace the $C$ in $C(1)$ with $B$, and the $D$ in $D(1)$ with $A$, we obtain the other non-LS event in this pair, $B(1)D(2)+C(2)A(1)$, which can be written as $B(1)D(2)+A(1)C(2)$ ($=S(1,2)+N(1,2)$).

\noindent
An alternative method for finding the second non-LS event in such a pair is to take the first non-LS event and switch the two elementary attributes that served as replacements. In the example given above, we would find the second event in the pair, namely, $S(1,2)+N(1,2)$ ($=B(1)D(2)+A(1)C(2)$), by switching the $A$ with the $B$ in the first non-LS event, $X(1,2)+Y(1,2)$ ($=A(1)D(2)+B(1)C(2)$). Each non-LS event in the pair is invariant with respect to a pair of simultaneous permutations of elementary attributes: permutation of the elementary attributes that served as replacements, and permutation of the remaining two elementary attributes. For example, each of the non-LS events in the pair given above is invariant with respect to the simultaneous permutations $(A,B)$ and $(C,D)$. Under these permutations, the non-LS event $X(1,2)+Y(1,2)$ ($=A(1)D(2)+B(1)C(2)$) becomes $B(1)C(2)+A(1)D(2)$ ($=Y(1,2)+X(1,2)$). To see that this is the original non-LS event ($X(1,2)+Y(1,2)$), simply switch the two terms. The invariance of the other non-LS event in that pair ($S(1,2)+N(1,2)$) with respect to the permutations $(A,B)$ and $(C,D)$ can be shown analogously. Every non-LS event is invariant with respect to a pair of simultaneous permutations, so every non-LS event is said to be {\it simultaneously symmetric} (SS). The ``$\cap$'' symbol is used to indicate simultaneous symmetry (e.g.,  $(A,B)\cap(C,D)$ denotes the simultaneous symmetry of each of the non-LS events in the example).

\noindent
The second pair of non-LS events generated from a given LS event can be found by applying the analogous procedure to the pair of elementary events with particle label 2. 
\end{enumerate}

\item  Since the combined attributes related to QI are $X=AD$ and $Y=BC$, the LS events naturally associated with QI are $X(1,2)+X(2,1)$ and $Y(1,2)+Y(2,1)$, which are regular events. The anti-events associated with QI are the four non-LS events which can be generated from either of these regular events. The combined attributes related to CI are $E=AB$ and $W=CD$, so the LS events naturally associated with CI are $E(1,2)+E(2,1)$ and $W(1,2)+W(2,1)$, which are anti-events. The regular events associated with CI are the four non-LS events which can be generated from either of these anti-events. Since we have posited the combined attributes related to RI as $N=AC$ and $S=BD$, the LS events associated with RI are $N(1,2)+N(2,1)$ and $S(1,2)+S(2,1)$, which are regular. The four non-LS events that can be generated from either of these regular events are $X(2,1)+Y(1,2),\ X(1,2)+Y(2,1),\ E(2,1)+W(2,1)$ and $E(1,2)+W(1,2)$. The former two are also regular events, and the latter two are the anti-events associated with RI.
\end{itemize}

In Table~\ref{TableClassification_p1} we list the regular events and anti-events associated with QI, CI, and RI, together with their symmetries. From the table, we can deduce a possibility for the role of the system RI within the overall scheme. 

\begin{table}[htbp]\caption{Classification of even events associated with QI, CI, and RI}
\scriptsize
\begin{tabular}{|c|c|c|}  \hline
&Regular events&Anti-events\\
&$<$Symmetries$>$&$<$Symmetries$>$\\ \hline
QI&$X(1,2)+X(2,1)=A(1)D(2)+A(2)D(1)$&$E(1,2)+W(2,1)=A(1)B(2)+C(2)D(1)$\\
&&$N(1,2)+S(2,1)=A(1)C(2)+B(2)D(1)$\\
&$<(A,D),\ (1,2)>$&$<(A,D)\cap(B,C)>$\\ \cline{2-3}
&$Y(1,2)+Y(2,1)=B(1)C(2)+B(2)C(1)$&$E(2,1)+W(1,2)=A(2)B(1)+C(1)D(2)$\\
&&$N(2,1)+S(1,2)=A(2)C(1)+B(1)D(2)$\\
&$<(B,C),\ (1,2)>$&$<(A,D)\cap(B,C)>$\\ \hline

CI&$N(1,2)+S(1,2)=A(1)C(2)+B(1)D(2)$&$E(1,2)+E(2,1)=A(1)B(2)+A(2)B(1)$\\
&$X(1,2)+Y(1,2)=A(1)D(2)+B(1)C(2)$&\\
&$<(A,B)\cap(C,D)>$&$<(A,B),\ (1,2)>$\\ \cline{2-3}
&$N(2,1)+S(2,1)=A(2)C(1)+B(2)D(1)$&$W(1,2)+W(2,1)=C(1)D(2)+C(2)D(1)$\\
&$X(2,1)+Y(2,1)=A(2)D(1)+B(2)C(1)$&\\
&$<(A,B)\cap(C,D)>$&$<(C,D),\ (1,2)>$\\ \hline

RI&$N(1,2)+N(2,1)=A(1)C(2)+A(2)C(1)$&$E(2,1)+W(2,1)=A(2)B(1)+C(2)D(1)$\\
  &$<(A,C),\ (1,2)>$&\\
&$X(1,2)+Y(2,1)=A(1)D(2)+B(2)C(1)$&\\
&$<(A,C)\cap(B,D)>$&$<(A,C)\cap(B,D)>$\\  \cline{2-3}
&$S(1,2)+S(2,1)=B(1)D(2)+B(2)D(1)$&$E(1,2)+W(1,2)=A(1)B(2)+C(1)D(2)$\\
&$<(B,D),\ (1,2)>$&\\
&$X(2,1)+Y(1,2)=A(2)D(1)+B(1)C(2)$&\\
&$<(A,C)\cap(B,D)>$&$<(A,C)\cap(B,D)>$\\ \hline
\end{tabular}\label{TableClassification_p1}\end{table}

\section{Concluding remarks}\label{SecRemarks_p1}
\begin{figure}[h]
\begin{center}
\includegraphics[width=130mm,height=74.6mm]{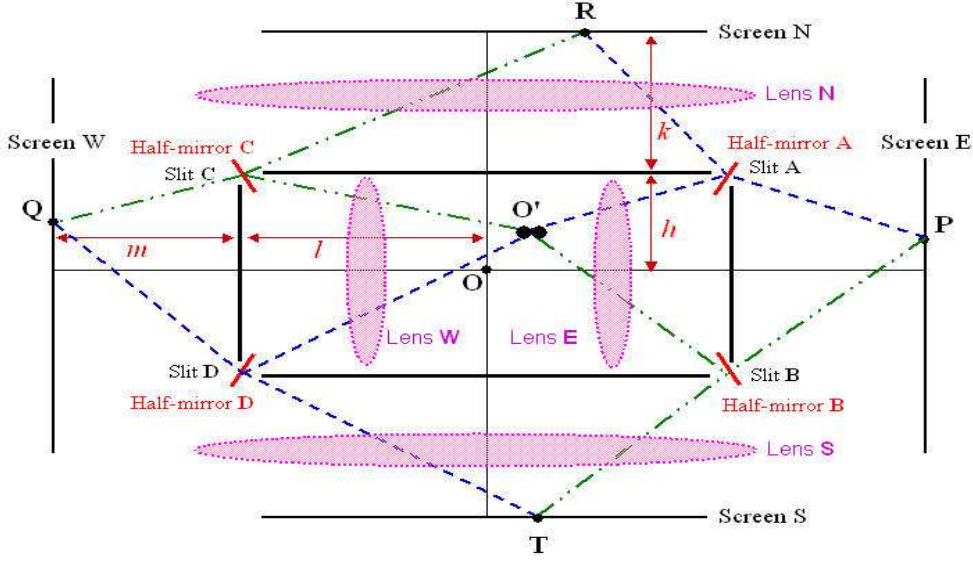}
\end{center}
\caption{An experiment to detect RI}\label{FigTwoDoubleSlit2_p1}
\end{figure}

We suggest an experiment to detect RI as shown in Fig.~\ref{FigTwoDoubleSlit2_p1}. The experimental configuration is basically same as the one in the two-double-slit experiment in Fig.~\ref{FigTwoDoubleSlit_p1}. The differences are that we we place two additional screens $\mathrm{N}$ and $\mathrm{S}$, four half-mirrors A, B, C, and D, and two additional walls on the lines between slit {\bf A} and slit {\bf C}, and between slit {\bf B} and slit {\bf D}. We may need to assume that $k$ is large compared to $l$ in a feasible experiment, so that, in addition, we place convex lens N, S, E, and W as in the figure. We will refer the experimental configuration in Fig.~\ref{FigTwoDoubleSlit_p1} to as the first configuration, and the one in Fig.~\ref{FigTwoDoubleSlit2_p1} to as the second configuration. 

A pair of photons are emitted at the origin $\mathbf{O}$. Similarly, there is some uncertainty in the coordinates of the point of emission, and the only small vertical displacement of point $\mathrm{O}^{\prime}$ from the origin is in consideration. Namely, one photon passes through slit $\bm{\mathrm{A}}$ and/or slit $\bm{\mathrm{B}}$, and reaches point $\bm{\mathrm{R}}$ on screen $\mathrm{N}$ or point $\bm{\mathrm{P}}$ on screen $\mathrm{E}$, and/or point $\bm{\mathrm{T}}$ on screen $\mathrm{S}$ or point $\bm{\mathrm{P}}$ on screen $\mathrm{E}$, and the other photon passes through slit $\bm{\mathrm{C}}$ and/or slit $\bm{\mathrm{D}}$, and reaches point $\bm{\mathrm{R}}$ on screen $\mathrm{N}$ or point $\bm{\mathrm{Q}}$ on screen $\mathrm{W}$ and/or point $\bm{\mathrm{T}}$ on screen $\mathrm{S}$ or point $\bm{\mathrm{Q}}$ on screen $\mathrm{W}$. 

We interpreted the combined event $X(1,2)=A(1)D(2)$, for example, as that in which photon 1 passes through slit $\bm{\mathrm{A}}$ and photon 2 passes through slit {\bf D}. Then we used such notations in Table~\ref{TableClassification_p1} and classified interference patterns in (\ref{Prob1_p1}) which is the probability amplitude that one photon reaches point $\mathbf{P}$ and the other photon reaches point $\mathbf{Q}$. 

We reinterpret the event $A(1)$ as that in which photon 1 passes through slit $\bm{\mathrm{A}}$ and then reaches either point $\mathbf{P}$ or point $\mathbf{R}$, and similarly for the other elementary events. Then, we can use Table~\ref{TableClassification_p1} as the classification of interference patterns in the second configuration as well.  

When $k\theta d\gg 1$ (where $\theta$ denotes the angle $\mathbf{AO^\prime B}$), QI occurs not only on screens E and W, but also on screens N and S. Whereas, when $k\theta d\ll 1$, CI occurs on screens E and W, and RI occurs on screens N and S. That is, if we rotated the diagram of the first configuration by 90 degrees, the expressions for the events in RI would remains unaffected, whereas the expressions for the events in CI would be switched with those for the events in RI. 

As shown in Table~\ref{TableClassification_p1}, the whole set of the regular events (resp. that of the anti-event) associated with CI are same as that of the regular events (resp. that of the anti-event) associated with RI. That is, the difference between CI and RI is about how to mix the combined events in each set. 

Let us express the event in which photon 1 reaches a point on screen $\mathrm{N}$ (resp. $\mathrm{S}$) as $U(1)$ (resp. $D(1)$), and similarly for $U(2)$ and $D(2)$. The set of all possible arrangements is:
\[\{U(1)U(2),\,D(1)D(2),\,U(1)D(2)+U(2)D(1)\}\]
This is same as that in Fig.~\ref{FigStatistics_p1}. Since two photons are emitted in clearly separated short intervals of time, we can separate only $U(1)U(2)$ and $D(1)D(2)$ from $U(1)D(2)+U(2)D(1)$ at the level of an elementary measurement. Namely, we can distinguish a pair of two even events associated with RI on a screen (i.e., either screen N or screen S) into an LS event and a non-LS event. CI does no have this feature. 

In this paper, we classified interference patterns in a two double-slit experiment in terms of symmetries of permutations of particle labels and attributes.  Our analysis suggests that RI is latent in CI, and permits us to change the combinations of combined events in CI.


\end{document}